\documentclass[fleqn,usenatbib]{mnras}
\usepackage{newtxtext,newtxmath}
\usepackage[T1]{fontenc}
\DeclareRobustCommand{\VAN}[3]{#2}
\let\VANthebibliography\thebibliography
\def\thebibliography{\DeclareRobustCommand{\VAN}[3]{##3}\VANthebibliography}
\usepackage{graphicx}
\usepackage{amsmath}
\usepackage{xspace}
\usepackage{multirow}

\title[GRB radio correlation]{Evidence for an intrinsic luminosity-decay correlation in GRB radio afterglows}

\author[S. P. R. Shilling et al.]{S. P. R. Shilling$^{1,2,3}$\thanks{E-mail: s.shilling@lancaster.ac.uk},
S. R. Oates$^{1}$,
D. A. Kann\thanks{Deceased}$^{4,5}$,
J. Patel$^{6}$,
J. L. Racusin$^{2}$,
B. Cenko$^{2}$,
R. Gupta$^{2,7}$,
\newauthor
M. Smith$^{1}$,
L. Rhodes$^{8}$,
K. R. Hinds$^{9}$,
M. Nicholl$^{10}$,
A. Breeveld$^{11}$,
M. Page$^{11}$,
M. De Pasquale$^{12}$,
\newauthor
and B. Gompertz$^{6}$
\\
$^{1}$ Department of Physics, Lancaster University, Lancaster, LA1 4YB, UK\\
$^{2}$ Astrophysics Science Division, NASA Goddard Space Flight Center, Mail Code 661, Greenbelt, MD, 20771, USA\\
$^{3}$ Center for Research and Exploration in Space Science and Technology, NASA Goddard Space Flight Center, Greenbelt, MD, 20771, USA\\
$^{4}$ Hessian Research Cluster ELEMENTS, Giersch Science Center, Max-von-Laue-Stra$\beta$e 12, Goethe University Frankfurt, Campus Riedberg, D-60438\\ Frankfurt am Main, Germany\\
$^{5}$ Instituto de Astrofísica de Andalucía (IAA-CSIC), Glorieta de la Astronomía s/n, 18008 Granada, Spain \\
$^{6}$ School of Physics and Astronomy \& Institute for Gravitational Wave Astronomy, University of Birmingham, B15 2TT, UK \\
$^{7}$NASA Postdoctoral Program Fellow \\
$^{8}$ Astrophysics, Department of Physics, University of Oxford, Denys Wilkinson Building, Keble Road, Oxford, OX1 3RH, UK \\
$^{9}$ Astrophysics Research Institute, Liverpool John Moores University, Liverpool Science Park, 146 Brownlow Hill, Liverpool L3 5RF, UK \\
$^{10}$ Astrophysics Research Centre, School of Mathematics and Physics, Queens University Belfast, Belfast BT7 1NN, UK \\
$^{11}$ University College London, Mullard Space Science Laboratory, Holmbury St. Mary, Dorking, RH5 6NT, UK\\
$^{12}$ University of Messina, MIFT Department, Polo Papardo, Viale F.S. D’Alcontres 31, 98166 Messina, Italy
}

\date{Accepted XXX. Received YYY; in original form ZZZ}

\pubyear{2025}

\begin{document}
\label{firstpage}
\pagerange{\pageref{firstpage}--\pageref{lastpage}}
\maketitle

\newcommand{\swift}{{\em Swift}\xspace}
\newcommand{\fermi}{{\em Fermi}\xspace}

\begin{abstract}
We present the discovery of a correlation, in a sample of 16 gamma-ray burst 8.5 GHz radio afterglows, between the intrinsic luminosity measured at 10 days in the rest frame, $L_{\mathrm{Radio,10d}}$, and the average rate of decay past this time, $\alpha_{>10d}$. The correlation has a Spearman's rank coefficient of $-0.70 \pm 0.13$ at a significance of $>3\sigma$ and a linear regression fit of $\alpha_{>10d} = -0.29^{+0.19}_{-0.28} \log \left(L_{\mathrm{Radio,10d}} \right) + 8.12^{+8.86}_{-5.88}$. This finding suggests that more luminous radio afterglows have higher average rates of decay than less luminous ones. We use a Monte Carlo simulation to show the correlation is not produced by chance or selection effects at a confidence level of $>3\sigma$. Previous studies found this relation in optical/UV, X-ray and GeV afterglow light curves, and we have now extended it to radio light curves. The Spearman's rank coefficients and the linear regression slopes for the correlation in each waveband
are all consistent within $1\sigma$. We discuss how these new results in the radio band support the effects of observer viewing geometry, and time-varying microphysical parameters, as possible causes of the correlation as suggested in previous works.

\end{abstract}


\section{Introduction}
\label{sec:introduction}
Gamma-ray Bursts (GRBs) are high-energy, short-lived and rapidly evolving transient astrophysical sources. Their emission is divided into two phases: prompt and afterglow. The prompt emission is the initial pulse of gamma-rays with a short lifetime, typically between a few milliseconds and a few thousand seconds \citep{mes97,lie16}. The afterglow emission is the associated broadband emission which is observable across the electromagnetic spectrum - from radio to X-ray and very high energies (VHE) - and has a longer observed lifetime, typically lasting hours to days \citep{sar98,eva09,Chandrafrail12}, and in some cases even weeks or years \citep[e.g. GRB 130427A and GRB 221009A;][respectively]{pasq16,rho24}. 

The standard fireball model offers an accepted explanation of the prompt emission and the afterglow, suggesting the GRB central engine releases a jetted ultrarelativistic `fireball' consisting of high energy photons with some degree of baryon and lepton loading \citep{Cavallo_78,pac86,goo86,she90,pac90,rees92,mes93}. In this model, the prompt emission is thought to be produced within the jet, mainly by non-thermal processes such as synchrotron and inverse Compton emission, when shells of ejecta with different Lorentz factors interact \citep{rees94,piran99,guetta01}. It is also thought that the prompt emission can in part be produced by photospheric emission since the optically thick fireball cools and eventually becomes transparent, enabling thermal (blackbody) emission \citep{goo86,pac86}. Also in this model, the afterglow is thought to be produced when the jet ploughs into the external medium, creating shocks and sweeping up material, which accelerates charged particles, producing synchrotron emission and synchrotron Self-compton emission \citep{rees92, mes93, mes97, sar98, sar01}. 

Previous studies in multiple wavebands have found a correlation between the luminosity measured in the early stages of the afterglow and the average rate of decay of the afterglow light curve past this time, implying that brighter afterglows decay more quickly on average. Originally, \cite{oates12} discovered this correlation in a sample of 48 optical/UV afterglows observed with the Ultraviolet/Optical Telescope \citep[henceforth UVOT;][]{roming} onboard NASA's Neil Gehrels \swift observatory \citep[henceforth \swift;][]{geh04} and calculated a Spearman's rank correlation coefficient, $R_\mathrm{sp}$, of $R_\mathrm{sp}=-0.58 \pm 0.11$ with a p-value of $1.90\times10^{-5}$. Next, \cite{Racusin16} found this correlation in a sample of 237 X-ray afterglows observed with \swift's X-ray Telescope \citep[henceforth XRT;][]{bur05} and calculated $R_\mathrm{sp} = -0.59 \pm 0.09$ with a p-value of $\ll10^{-5}$. Recently, \cite{Hinds23} found this correlation in a sample of 13 GeV afterglows observed with the Large Area Telescope (henceforth LAT) onboard the \fermi Gamma-ray Space Telescope \citep[henceforth \fermi;][]{atwood09} and calculated $R_\mathrm{sp} = -0.74 \pm 0.19$ with a p-value of $4.11\times10^{-3}$. The strength of the correlation is consistent between these wavebands within $1\sigma$ \citep{oates15,Hinds23}. 

In this paper, we build upon these studies of the luminosity-decay correlation by extending the analyses to a sample of GRB afterglow light curves in the radio band, at 8.5 GHz, thereby covering 15 orders of magnitude in frequency (from GeV photons to radio photons). The analysis, including sample selection, is given in \S\ref{sec:data}; the results of the correlation analysis are given in \S\ref{sec:results}; a discussion of the results - covering a multiwavelength comparison, the possible physical causes of the correlation, and its application to different studies - is given in \S\ref{sec:discussion}; and the conclusions are given in \S\ref{sec:conclusion}. For this study, we assume cosmological parameter values of $H_0 = 70$ km s$^{-1}$ Mpc$^{-1}$, $\Omega_{\Lambda} = 0.7$ and $\Omega_{m} = 0.3$, and the flux convention of $F(t,\nu) \propto t^{\alpha} \nu^{\beta}$, where $\alpha$ and $\beta$ are the temporal and spectral indices, respectively.

\section{Data analysis}
\label{sec:data}

\subsection{Sample selection}
\label{sec:sample}
We use a catalogue of GRB radio observations that were compiled by D. A. Kann from the literature published between 1997 and 2020. We supplemented this with observations reported in the literature from 2020 up until February 2024. Our catalogue contains 480 GRBs consisting of 6227 flux density measurements, including upper limits. The data spans a frequency range of 0.4 - 667 GHz, with the majority ($\sim25\%$) concentrated around 8.5 GHz. Within our catalogue, 55\% of the data were collected with the Very Large Array \citep[VLA;][]{thompson80}, 18\% with the Arcminute Microkelvin Imager – Large Array \citep[AMI-LA;][]{zwart08}, 8\% with the Australia Telescope Compact Array \citep[ATCA;][]{frater92} and the remaining 19\% from a mixture of 40 other radio telescopes. 

\begin{figure}
    \includegraphics[width=\columnwidth]{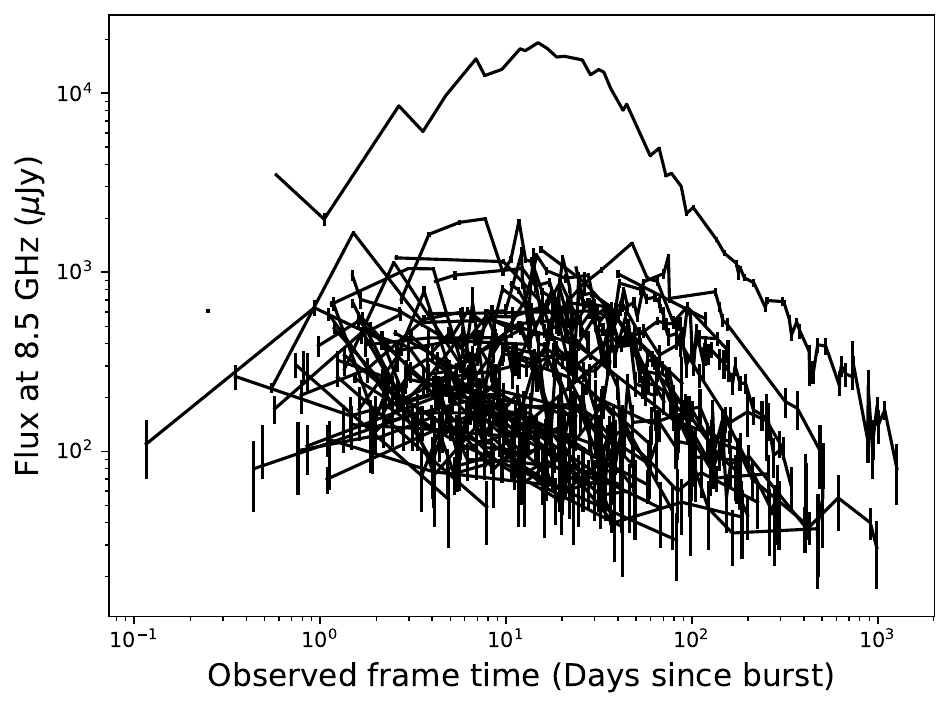}
    \caption{The flux density radio light curves at 8.5 GHz for the 81 GRBs with a measured redshift and a peak flux of $> 100 \mathrm{\ \mu Jy}$. For clarity, we only plot measurements where SNR $>2$.}
    \label{fig:initial distribution}
\end{figure}

We select the sample from the catalogue according to the following specific criteria. Firstly, we select the 246 GRBs which have a measured redshift. This is required in order to convert from flux density to rest frame luminosity (see \S\ref{sec:light curves}). Secondly, we exclude upper limits in order to make error-weighted power-law fits (see \S\ref{sec:measurements}). This reduces the number of GRBs from 246 to 207. Thirdly, we select data within a narrow frequency range of 8.5 $\pm$ 0.1 GHz. This frequency range contains the largest number of individual GRBs (131) and the best sampled light curves, akin to the case in \cite{Chandrafrail12}. As such, our sample is further reduced to 131 GRBs. Finally, we require light curves to have a peak flux of $> 100 \mathrm{\ \mu Jy}$ to ensure sufficient signal-to-noise and sampling. Of the 131 GRBs in our sample thus far, we select 81 of which that satisfy this final criterion. Figure \ref{fig:initial distribution} shows the observed frame radio light curve distribution at 8.5 GHz for these 81 selected GRBs.

\subsection{Light curves}
\label{sec:light curves}
We convert the time of each measurement in each light curve into the rest frame using:
\begin{equation}
\label{equation:times}
t_{\mathrm{rest}}=\frac{t_\mathrm{obs}}{1+z},
\end{equation}
where ${t_\mathrm{obs}}$ is the observed time elapsed since the start of the GRB, $t_{\mathrm{rest}}$ is the rest frame time, and $z$ is the GRB's redshift. The light curves are converted from flux density to intrinsic luminosity and k-corrected to 8.5 GHz in the co-moving frame using:
\begin{equation}
\label{equation:luminosity}
L_{\nu}=4 \pi F_{\nu} d_{L}^2 (1+z)^{-\beta -1},
\end{equation}
where $L_{\nu}$ is the luminosity in erg s$^{-1}$ Hz$^{-1}$, $F_{\nu}$ is the observed flux density, and $d_{L}$ is the luminosity distance \citep{blo01}. We do not know $\beta$ for each GRB, so we assume a single value to use for all GRBs using the standard fireball model. Accordingly, we assume the slow cooling scenario for an adiabatic fireball expanding into a uniform medium, consistent with what has been found for GRB afterglows \citep{sch11,gom18}. As in the other wavebands, we are testing for the correlation during the decaying phase of the light curves. Therefore, we assume that the synchrotron frequencies are ordered as $\nu_{m} <\nu < \nu_c$, where $\nu_{m}$ is the peak frequency and $\nu_c$ is the cooling frequency \citep{sar98}. This corresponds to when light curves are expected to be in the early decay phase. As such, we use a value of $\beta = -(p-1)/2$ for all light curves in our sample, where $p$ is the electron energy index \citep{sar98} and is assumed to be $p=2.36$ \citep{cur10}. 

\begin{figure}
    \includegraphics[width=\columnwidth]{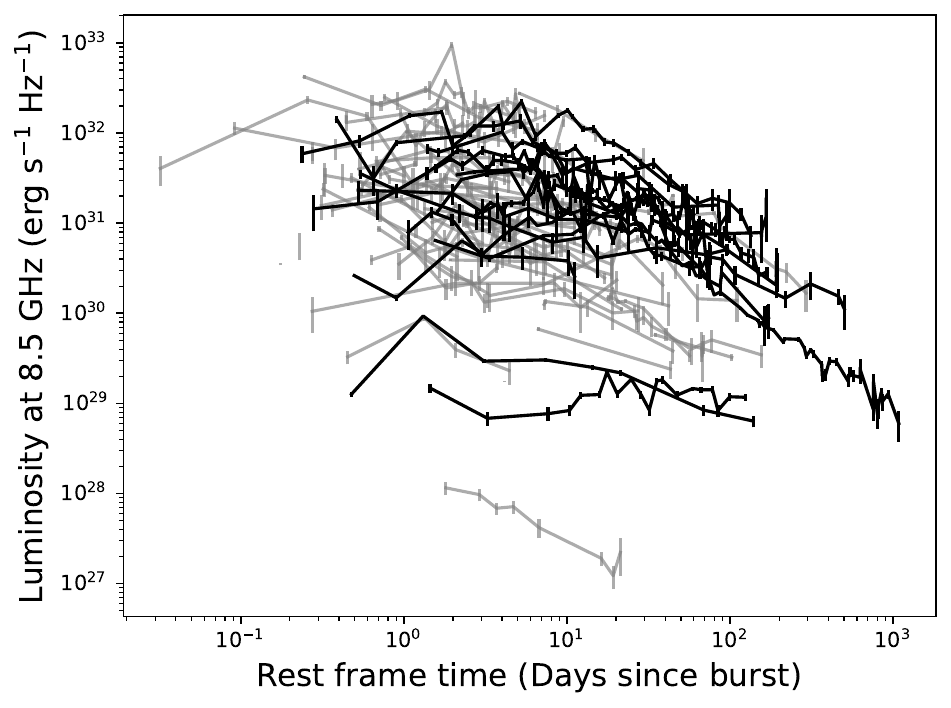}
    \caption{The rest frame radio light curves at 8.5 GHz for the 81 GRBs in our initial sample in grey, with the light curves of the final selected 16 GRBs overlaid in black. Measurements with SNR $<2$ are excluded in this figure.}
    \label{fig:final_sample}
\end{figure}

\subsection{Determining the luminosity and average rate of decay}
\label{sec:measurements} 
First, we choose the time at which we measure the luminosity and the average rate of decay from. This time must be early enough so that the light curves are still sufficiently bright, but not too early as to avoid the rising phase. Motivated by the average light curve peak time of 5.5 days in the rest frame in our sample of 81 selected light curves, we choose $10$ days in the rest frame for this time. We therefore measure the radio luminosity at 10 days, $L_{\mathrm{Radio,10d}}$ and average rate of decay from 10 days until the end of observations, $\alpha_{>10d}$, using two separate power-laws of the form:
\begin{equation}
\label{equation:powerlaweqn}
L=Nt^{\alpha},
\end{equation}
where $N$ and ${\alpha}$ are the normalization and the index of the power-law. 

To constrain a single power-law fit, three or more data points are required. As such, we define two time ranges (one for each measurement) to fit independently with a single power-law to each. For the average rate of decay, this time range is simply $\geq10$ days. For the luminosity at 10 days, we define a time range of $\pm 0.4$ dex (centered on 10 days). This time range optimizes the number of light curves containing three or more data points in this region while being small enough to avoid contamination from the rising phase. Out of the 81 GRBs in our sample, 52 do not have three or more data points in at least one of the two time ranges (of $10^{1 \pm 0.4}$ days or $\geq 10$ days). We therefore exclude these GRBs from further analysis as the power-law cannot be reasonably constrained for at least one of the two fits. This reduced our sample from 81 GRBs to 29. 

For each of these 29 GRBs, we fit a single power-law to the data in each of the time ranges (of $10^{1 \pm 0.4}$ days and $\geq 10$ days) using the \texttt{lmfit} module in \texttt{python}. We then measure $L_{\mathrm{Radio,10d}}$ and $\alpha_{>10d}$ for each light curve. We measure $L_{\mathrm{Radio,10d}}$ by evaluating Equation \ref{equation:powerlaweqn} at $t=10$ days, using the best fitting values of $N$ and $\alpha$ for the power-law fit to the data within $10^{1 \pm 0.4}$ days and convert this measurement into log space. Additionally, $\alpha_{>10d}$ is given by the best fitting value of $\alpha$ for the power-law fit to the data at $\geq10$ days. We note that individual light curves may have unique features such as; plateaus, forward shock decays, and post jet-break decays \citep{nousek}. However, we use a single power-law when measuring $\alpha_{>10d}$ regardless of any unique features that an individual light curve may have, as we are only interested in the average rate of decay.

\begin{figure}
    \includegraphics[width=\columnwidth]{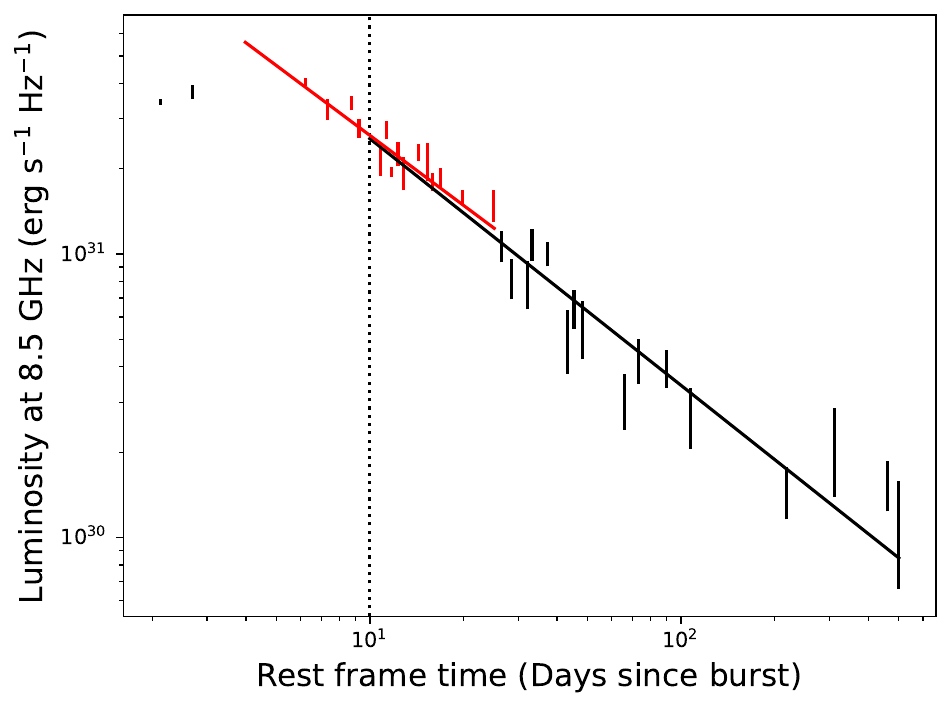}
    \caption{The 8.5 GHz light curve of GRB 980703A, one of the GRBs in our final sample of 16. The vertical dotted line marks 10 days. The red points are those within $10^{1 \pm 0.4}$ days and the solid red line is the fit to these points. The black points are the rest of the light curve. The solid black line is the fit to the points $\geq$ 10 days, which overlaps the red points from 10 days up to $10^{1 + 0.4}$ days.}
    \label{fig:light curve example}
\end{figure}

We only consider light curves which have well-constrained measurements, and therefore select those which have an uncertainty of $\leq 0.5$ on their measurements of $\alpha_{>10d}$ and $\mathrm{log_{10}}\left(L_{\mathrm{Radio,10d}}\right)$. This reduced the sample from 29 GRBs to 16, constituting our final sample. Figure \ref{fig:final_sample} shows the distribution of rest frame light curves in the final sample of 16 GRBs overlaid onto the distribution of the initial sample of 81. Figure \ref{fig:light curve example} shows an example of an individual rest frame light curve, GRB 980703A, with the two single power-law fits overlaid. Table \ref{tab:measurements} summarizes the parameters measured for each GRB in the final sample. All of the GRBs in our final sample of 16 are classified as long GRBs (LGRBs) based on their $T_{90}$ values, the time over which 90\% of the prompt emission is observed, exceeding 2 seconds \citep[for the $T_{90}$ values, see;][]{Chandrafrail12,Ashall19,abd09}.

\begin{table}
	\centering
	\caption{Parameters derived for the 16 GRBs in the final 8.5 GHz sample. Col. (1): GRB name. Col. (2): Redshift. Col. (3): Measured average rate of decay $\geq$ 10 days. Col. (4): Log of the measured luminosity at 10 days.}
	\label{tab:measurements}
	\begin{tabular}{lcccc}
	\hline
	GRB & Redshift  & $\alpha_{>10d}$ & $\log_{10}\left(L_{\mathrm{Radio,10d}}\right)$\\
	& & &[erg s$^{-1}$ Hz$^{-1}$] \\
	(1) & (2) & (3) & (4) \\
	\hline
    970508 & 0.83 &-0.55 $\pm$ 0.05 & 31.26 $\pm$ 0.09 \\
    980703 & 0.97 & -0.87 $\pm$ 0.05 & 31.42 $\pm$ 0.08 \\
    991208 & 0.71 & -0.87 $\pm$ 0.08 & 31.21 $\pm$ 0.11 \\	
    000301C & 2.04 & -1.06 $\pm$ 0.13 & 31.75 $\pm$ 0.07 \\
    000418 &1.12 &-1.27 $\pm$ 0.08 & 31.61 $\pm$ 0.15 \\
    000911 & 1.06 &-0.60 $\pm$ 0.40 & 30.40 $\pm$ 0.43 \\
    000926 & 2.04 &-0.77 $\pm$ 0.17 & 31.71 $\pm$ 0.22 \\
    010222 & 1.48 &-0.45 $\pm$ 0.20 & 30.73 $\pm$ 0.19 \\
    021004 & 2.33 &-1.32 $\pm$ 0.05 & 32.07 $\pm$ 0.18 \\
    030329 & 0.17 &-0.93 $\pm$ 0.04 & 31.03 $\pm$ 0.09 \\
    031203 & 0.11 & -0.03 $\pm$ 0.09 & 28.98 $\pm$ 0.37 \\
    070125 & 1.55 &-0.78 $\pm$ 0.09 & 31.78 $\pm$ 0.12 \\
    090902B & 1.82 & -0.58 $\pm$ 0.44 & 30.88 $\pm$ 0.23 \\
    091020 & 1.71 &-0.64 $\pm$ 0.27 & 31.37 $\pm$ 0.11 \\
    100418A & 0.62 & 0.09 $\pm$ 0.24 & 30.91 $\pm$ 0.22 \\
    161219B & 0.15 & -0.65 $\pm$ 0.07 & 29.44 $\pm$ 0.01 \\
    \hline
	\end{tabular}
    \begin{minipage}[c]{\columnwidth}
    {{\it Notes.} \scriptsize GRB redshift references -- 970508 \citep{blo98}; 980703 \citep{djo98}; 991208 \citep{cas01}; 000301C \citep{jen01}; 000418 \citep{blo03b}; 000911 \citep{pri02}; 000926 \citep{cas03}; 010222 \citep{mir02}; 021004 \citep{cas10}; 030329 \citep{tho07}; 031203 \citep{mar07}; 070125 \citep{cia11}; 090902B \citep{cen11}; 091020 \citep{xu09}; 100418A \citep{pos18}; 161219B \citep{tan16,can17}.}
    \end{minipage}
\end{table}

\subsection{Correlation analysis}
\label{sec:correlation}
We perform a Spearman's rank test between the measurements of $L_{\mathrm{Radio,10d}}$ and $\alpha_{>10d}$ to determine if these parameters are correlated. This is calculated using the \texttt{scipy.stats} module in \texttt{python} \citep{scipy20}. We also perform a partial Spearman's rank test between these measurements of $L_{\mathrm{Radio,10d}}$ and $\alpha_{>10d}$, wherein the redshift is treated as a confounding variable \citep{ken79}. This determines the strength of the correlation after removing the effect of redshift, and is calculated using the \texttt{Pingouin} module in \texttt{python} \citep{vallat18}. 

We perform an error-weighted linear regression analysis of the measurements of $L_{\mathrm{Radio,10d}}$ and $\alpha_{>10d}$ to characterize the relationship between these variables. Linear regression provides the least squares best fitting parameters for the slope and intercept. The linear regression is calculated using the \texttt{scipy.odr} module in \texttt{python} which has the advantage of weighting the calculation by the errors in both variables \citep{scipy20}.

Errors on the Spearman's rank coefficient and the linear regression parameters are determined via Monte Carlo Bootstrap calculations due to the small sample size \citep[<50;][]{iso90, fei92}. We randomly resample, with replacement, the measurements of $L_{\mathrm{Radio,10d}}$ and $\alpha_{>10d}$ in pairs for $10^{5}$ trials. For each trial, we calculate the Spearman's rank coefficient and the linear regression parameters and record these simulated results. After $10^{5}$ trials, the distributions of each simulated result are ordered by ascending value. For each result, the $\pm1\sigma$ lower and upper errors are calculated by taking the difference between the mean of the simulated results and the 15.9th (lower) and the 84.1st (upper) percentile values of the simulated results, respectively.

\section{Results}
\label{sec:results}

\subsection{Luminosity-decay correlation}
\label{sec:correlation_sub_section}
\begin{center}
\begin{figure}
	\includegraphics[width=\columnwidth]{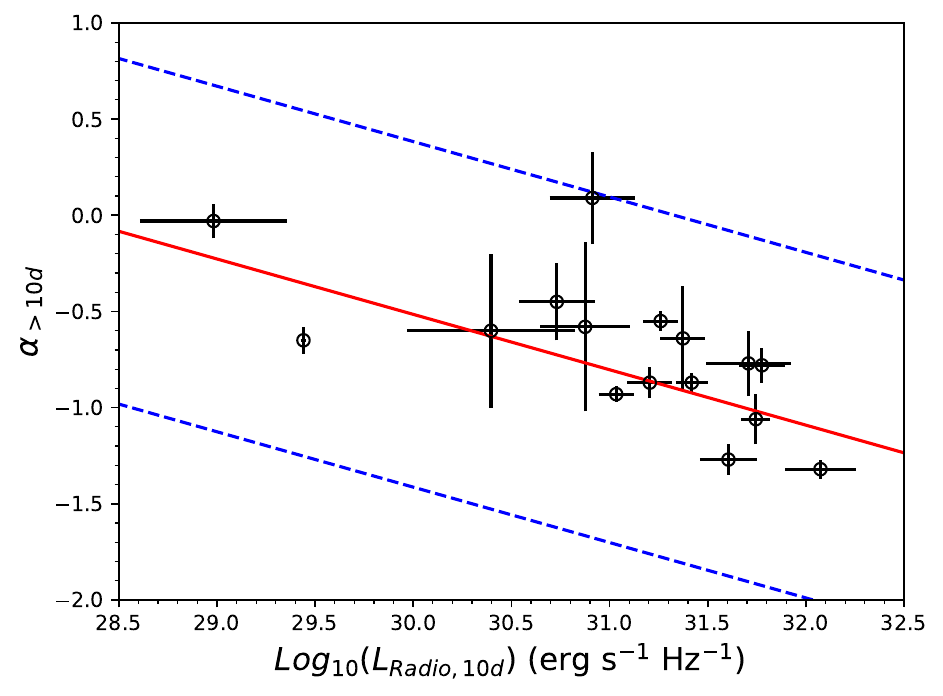}
    \caption{The average rate of decay from 10 days, $\alpha_{>10d}$, against the log of the radio luminosity at 10 days, $L_{\mathrm{Radio,10d}}$, for the 16 GRBs in the final 8.5 GHz sample. The red line is the best fit linear regression and the dashed blue lines are the $3\sigma$ (root-mean-square) deviation.}
    \label{fig:Correlation}
\end{figure}
\end{center}

It is apparent in Figure \ref{fig:final_sample} that the distribution of luminosity is widest at early times and becomes narrower at late times, suggesting the presence of the correlation. Figure \ref{fig:Correlation} shows a plot of the average rate of decay against the luminosity for each of the 16 light curves in the final sample with the linear regression fit overlaid. We calculate a Spearman's rank coefficient of $R_\mathrm{sp}=-0.70 \pm 0.13$ at a significance of $>3 \sigma$, indicating a strong negative correlation between the intrinsic luminosity at 10 days in the rest frame and the average rate of decay past 10 days in the rest frame, implying that more luminous radio afterglows decay more quickly on average than less luminous ones. To account for any dependence of the correlation parameters on redshift, we conduct a partial Spearman's rank test wherein the redshift is treated as a confounding variable. We calculate a partial Spearman's rank correlation coefficient of $-0.65$ with a p-value of $8.90 \times 10^{-3}$. This suggests that redshift does not cause the correlation. We calculate a linear regression slope and intercept of $-0.29^{+0.19}_{-0.28}$ and $8.12^{+8.86}_{-5.88}$, respectively. 

\subsection{Testing biases and assumptions}
\label{sec:biases}

\subsubsection{Selection criteria and randomness}
We examine the possibility that the observed correlation could be due to chance, or be artificially produced by our selection effects. Specifically, the selection of afterglows with peak fluxes greater than $100$ $\mu\mathrm{Jy}$. To address this and rule out these possibilities, we run a Monte Carlo simulation in which we evaluate a single power-law from 10 days (representing a synthetic light curve) using randomly selected parameters, determine whether they meet our selection criteria based on the synthetic peak flux, and test for the correlation amongst those that do. Although the average peak time of the light curves in our sample is 5.5 days, the light curves tend to have a slow turnover in their transition from the rising phase to the decay phase (as seen most clearly in Figure \ref{fig:final_sample}) meaning that the flux at 10 days still remains relatively similar to the peak flux. Therefore, the simulated flux of a single power-law plotted from 10 days in our simulation approximately represents the peak fluxes as used in our selection criteria.

For $10^6$ trials, we randomly select (with replacement) a value of $L_{\mathrm{Radio,10d}}$, $\alpha_{>10d}$, and redshift, from the uniform distributions of each parameter between the ranges of values in our sample. We randomly select a value of 
$L_{\mathrm{Radio,10d}}$ and $\alpha_{>10d}$ to derive a synthetic rest frame light curve by evaluating a single power-law (see Equation \ref{equation:powerlaweqn}). The simulated light curve is converted to the observed frame using the randomly selected redshift, assuming the same value of $\beta$ as in \S\ref{sec:light curves}. We test if the synthetic observed frame light curve meets our selection criteria and reject any light curves which do not. We store the original values of $L_{\mathrm{Radio,10d}}$ and $\alpha_{>10d}$ for those that do. For each trial, this process is repeated until there are 16 light curves which meet our selection criteria. We then test for a correlation between the 16 corresponding pairs of $L_{\mathrm{Radio,10d}}$ and $\alpha_{>10d}$ by running a Spearman's rank test. Out of $10^{6}$ trials, only 0.17\% of the simulated samples have Spearman's rank coefficients equal to or more negative than that of the real observed correlation. This suggests that, at a confidence level of $> 3 \sigma$, the correlation is intrinsic and is not produced by our selection criteria or by chance.

\begin{center}
\begin{figure}
	\includegraphics[width=\columnwidth]{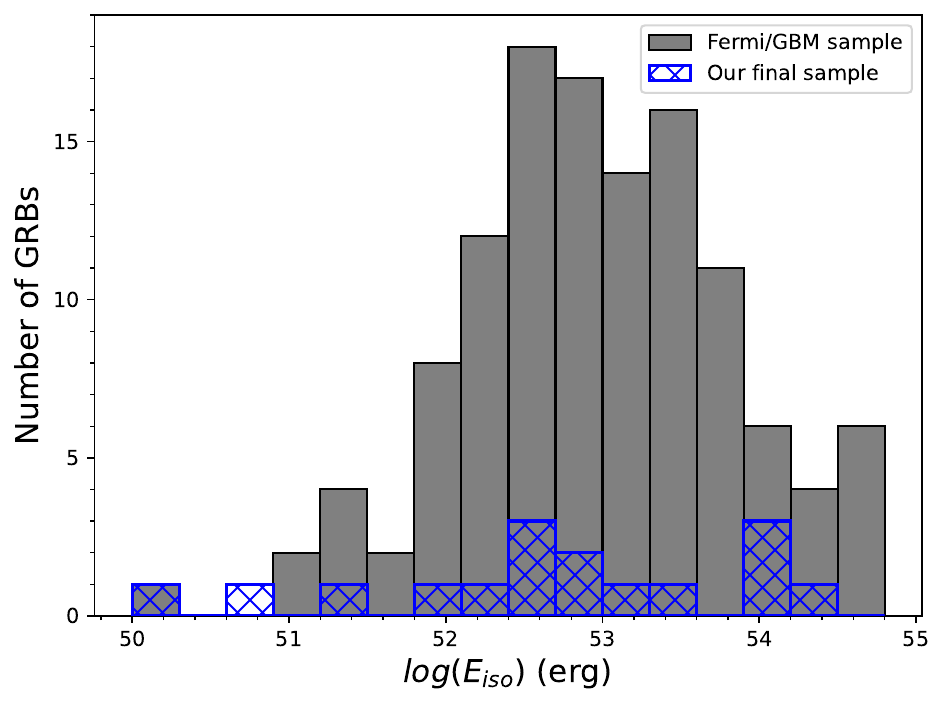}
    \caption{Log distribution of the isotropic $\gamma$-ray energy, $E_\mathrm{iso}$, of the 16 GRBs in our sample (cross hatched blue) and 121 GRBs observed with the \fermi/GBM \protect\citep[grey;][]{poolakkil21}. With a KS statistic of 0.15, our sample is not statistically different from the wider population of LGRBs observed with the \fermi/GBM.}
    \label{fig:eisodistcomparison}
\end{figure}
\end{center}

\subsubsection{Isotropic $\gamma$-ray energy bias}
\label{sec:eiso_dist_comparison}
It is a possibility that the LGRBs in our final sample are not representative of the wider LGRB population. For instance, we select those which are well sampled and have a measured redshift. These may introduce a bias towards brighter GRBs, as it is easier to obtain a spectrum for brighter afterglows. Consequently, we test how the distribution of GRB isotropic $\gamma$-ray energy, $E_\mathrm{iso}$, in our final sample compares to that of a wider sample of LGRBs. We use $E_\mathrm{iso}$ values measured in the 1-10,000 keV range presented in \cite{Chandrafrail12} for our entire sample apart from that of GRB 161219B, which was retrieved from \cite{Ashall19}. For the wider sample of LGRBs, we use $E_\mathrm{iso}$ values measured in the 1-10,000 keV range (using the Band spectral model) by \fermi's Gamma-ray Burst Monitor (GBM) for 121 LGRBs \citep{poolakkil21}. Figure \ref{fig:eisodistcomparison} shows these distributions. Additionally, we use a two sample Kolmogorov-Smirnov (KS) test to quantify how these two samples differ. We calculate a KS statistic of 0.15 and a p-value of 0.85, implying that the two distributions of $E_\mathrm{iso}$ are indeed drawn from the same population and, therefore, that our selected sample is not biased towards brighter GRBs. However, we note that the $E_\mathrm{iso}$ values for the GRBs in our final sample are measured by a range of instruments, and therefore may have different band passes and sensitivities compared to \fermi.

\begin{center}
\begin{figure}
	\includegraphics[width=\columnwidth]{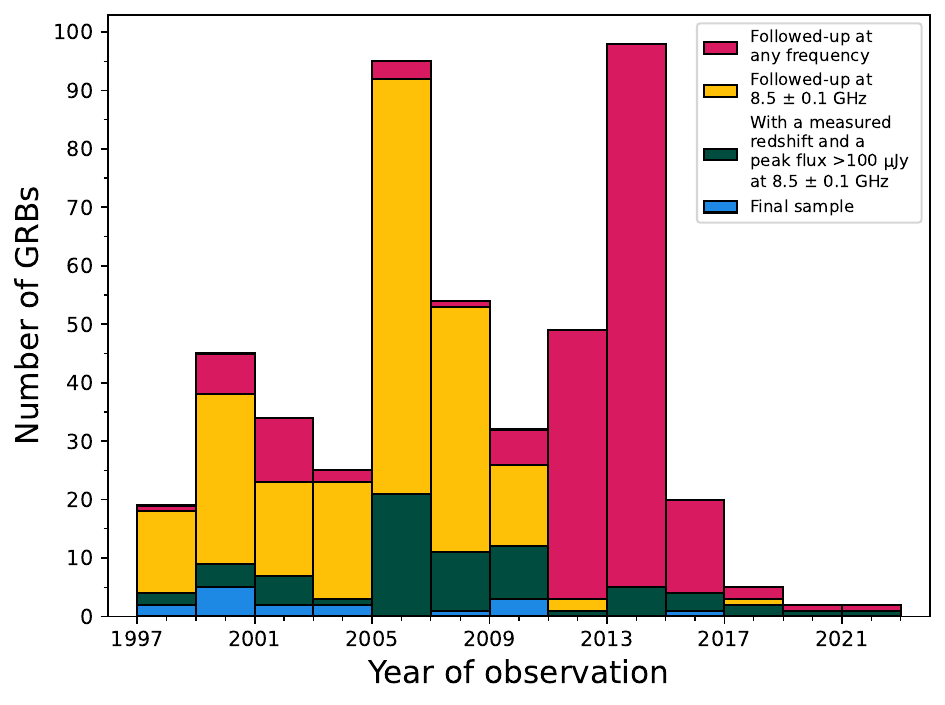}
    \caption{The distribution of GRBs with radio observations per year (using a bin size of two years). Pink shows the number of GRBs in our catalogue with radio follow-up, at any frequency. Yellow shows the number of GRBs which were observed at $8.5 \pm 0.1$ GHz. Green shows the number of GRBs observed at $8.5 \pm 0.1$ GHz which also have measured redshift and a peak flux $> 100 \mathrm{\ \mu Jy}$. Blue shows our final sample of 16 GRBs.}
    \label{fig:histogram_by_year}
\end{figure}
\end{center}

\subsubsection{Year of detection}
\label{sec:year}
Our final sample mostly consists of pre-\swift GRBs (up to 2005), with only 1 GRB after 2011. Figure \ref{fig:histogram_by_year} shows the yearly distribution of GRBs in our catalogue for those observed at any frequency and those observed at $8.5 \pm 0.1$ GHz, also for those observed $8.5 \pm 0.1$ GHz which have a measured redshift and a peak flux of $> 100 \mathrm{\ \mu Jy}$ (see \S\ref{sec:sample}), and for our final sample. Firstly, Figure \ref{fig:histogram_by_year} indicates that GRBs are mainly followed-up at $8.5 \pm 0.1$ GHz, up until 2011 after which different frequency bands are used. This may be a contributing factor as to why there is a lack of post-2011 GRBs in our final sample. Secondly, Figure \ref{fig:histogram_by_year} shows that, out of those observed at $8.5 \pm 0.1$ GHz with a measured redshift and a peak flux of $> 100 \mathrm{\ \mu Jy}$, a relatively large fraction are selected for our final sample in the pre-\swift era (up to $\sim 2005$) compared to in the early post-\swift era (between $\sim 2005-2011$). Since we select GRBs for our final sample based on how well sampled their light curves are, we interpret this difference as being due to post-\swift GRBs not being followed-up as extensively at $8.5 \pm 0.1$ GHz as pre-\swift GRBs. This is another probable factor explaining why there are relatively few post-\swift GRBs (prior to 2011) in our final sample. Finally, the number of GRBs with radio follow-up at any frequency is low from $\sim2017$ onwards. This may be due to reduced follow-up, or more likely that the data for these recent GRBs has not yet been published.

We further investigated the absence of post-2011 GRBs from our final sample by looking at the light curves of notable examples, including GRBs 130427A, 130907A, 171205A, 190114C, and 221009A. These GRBs make our initial cuts of having a measured redshift and a peak flux of $> 100 \mathrm{\ \mu Jy}$ at $8.5 \pm 0.1$ GHz but they are excluded from our final sample because they have less than 3 data points at 8.5 $\pm$ 0.1 GHz in at least one of the two time ranges (of $10^{1 \pm 0.4}$ days or $\geq 10$ days; see \S\ref{sec:measurements}).

\subsubsection{Frequency range}
As mentioned in \S\ref{sec:year}, there are typically fewer GRBs observed at $8.5 \pm 0.1$ GHz per year after 2011. Furthermore, some of the notable GRBs mentioned in \S\ref{sec:year} are well sampled at different frequencies, such as 130427A at 4.8 GHz, 190114C at 97.5 GHz, and GRB 221009A at 15.5 GHz. While others have observations covering a large range in frequency, but only a few data points per frequency, such as; 130907A from 5.0 to 24.5 GHz, and 171205A from 0.4 to 44 GHz. Therefore, by increasing our frequency range, we may be able to increase our sample size, and in particular, include a larger number of GRBs observed post-2011.

Our primary reason for initially selecting data within a narrow frequency range of 8.5 $\pm$ 0.1 GHz was because afterglows exhibit significant colour evolution around the peak. This is potentially due to the passage of the peak synchrotron frequency, $\nu_m$, through the radio bands, implying that the peak luminosity, the peak time, and the time of the onset of the decay phase will be different in different radio bands. Since we measure the luminosity soon after the light curves peak, constructing our light curves using a wider range of frequencies instead of a single frequency would affect the measurements of the luminosity at 10 days, and increase scatter in our results. Disregarding the possible effects of colour evolution, we increase the frequency range used to construct light curves to 8.5 $\pm$ 1.0 GHz, and to 8.5 $\pm$ 8.0 GHz, and repeat the analysis. These wider frequency ranges result in samples of 23 GRBs and 29 GRBs (including 5 and 9 GRBs after 2011) respectively. For the 8.5 $\pm$ 1.0 GHz, and 8.5 $\pm$ 8.0 GHz frequency ranges, we calculate Spearman's rank coefficients of $R_\mathrm{sp}=-0.14 \pm 0.26$ and $R_\mathrm{sp}=-0.14 \pm 0.20$ respectively, as well as p-values of 0.53 and 0.47 respectively.

\subsubsection{Spectral assumptions}
When k-correcting and converting the light curves to the rest frame in \S\ref{sec:light curves}, we assumed a spectral index for all light curves in the sample. Since the light curves are decaying and we are considering a slow cooling environment, we assume a spectral regime of $\nu_{m} < \nu < \nu_{c}$ and thus $\beta = -(p-1)/2$ \citep{sar98}. However, for the light curve k-correction and rest frame conversion in \cite{Chandrafrail12} the spectral regime of $ \nu < \nu_{m} < \nu_{c}$ was assumed, which corresponds to a value of $\beta = 1/3$ and is representative of the rising phase before $\nu_m$ has passsed through the radio band, instead of the early decay phase. To test the impact of this different choice of spectral regime, we repeat the analysis but instead assume $ \nu < \nu_{m}$, which corresponds to $\beta = 1/3$ as in \cite{Chandrafrail12}, and compare the difference. 

This change in the value of $\beta$ caused a negligible difference in the results, the most noticeable of which are that the average measurement of $L_{\mathrm{Radio,10d}}$ decreases from $2.8 \times 10^{31}$ erg s$^{-1}$ Hz$^{-1}$ to $1.1 \times 10^{31}$ erg s$^{-1}$ Hz$^{-1}$, and the range in $L_{\mathrm{Radio,10d}}$ decreases from $\sim3.1$ dex to $\sim2.6$ dex. Using $\beta =1/3$, we calculated a Spearman's rank coefficient of $R_\mathrm{sp}=-0.74 \pm 0.13$, a p-value of $1.01\times10^{-3}$, and a linear regression slope of $-0.34^{+0.29}_{-0.45}$.

As previously mentioned, we assumed the slow-cooling scenario for our analysis. This is consistent with the findings of \cite{lev23} which suggests the slow-cooling scenario is favoured over the fast-cooling scenario for a sample of radio afterglows that were found to have a clear break and to be consistent with the standard fireball model. Nevertheless, we test how our assumption of slow-cooling may affect the correlation (despite being the expected scenario), by repeating the analysis, assuming the fast-cooling scenario instead. In this case, we use a value of $\beta=-1/2$ corresponding to the equivalent of the early decay phase in this scenario; $ \nu_{c} < \nu < \nu_{m}$ \citep{sar98}, and calculate a Spearman's rank coefficient of $R_\mathrm{sp}=-0.70 \pm 0.13$, a p-value of $2.58\times10^{-3}$, and a linear regression slope of $-0.30^{+0.21}_{-0.32}$. These results are consistent with those we find for the slow-cooling scenario, where $\beta = -(p-1)/2$, indicating that the correlation is unlikely to be affected by assuming a different cooling scenario or spectral regime.

\subsubsection{Measurement time}
In \S\ref{sec:measurements} we chose 10 days in the rest frame as the optimal time to measure the luminosity and the time from which we measure the average decay index. We tested whether this time is optimal by re-measuring the parameters at later times of 15, 20 and 25 days and performing the correlation analysis for comparison. The Spearman's rank values calculated at later times were weaker, ranging between $ -0.3 \lesssim R_\mathrm{sp} \lesssim -0.25$. Furthermore, the distribution of measured luminosity decreases with each successive later time, with a decrease in range from $\sim3.1$ dex to $\sim2.6$ dex and in standard deviation from $\sim0.8$ dex to $\sim0.6$ dex. This decrease in luminosity dispersion at later times can be seen graphically in Figure \ref{fig:final_sample} and suggests the presence of the correlation, as mentioned in \S\ref{sec:correlation_sub_section}. On this basis, we established that 10 days was the optimal time without being too late as to decrease the range in measured luminosity. We could not go earlier than 10 days as the light curves are rising.

\subsubsection{Measurement uncertainty}
As detailed in \S\ref{sec:measurements}, our final selection criterion is to select light curves which have an uncertainty of $\leq 0.5$ on their measurements of $\alpha_{>10d}$ and $\mathrm{log_{10}}\left(L_{\mathrm{Radio,10d}}\right)$. We test the impact of this selection criterion by relaxing this constraint and repeating the analysis. Using larger thresholds of 0.75 and 1.0, we find the correlation is still observed but with slightly weaker Spearman's rank coefficients of $R_\mathrm{sp}=-0.57\pm 0.16$ and $R_\mathrm{sp}=-0.62 \pm 0.11$ respectively, and p-values of $1.06 \times 10^{-2}$ and $1.52 \times 10^{-3}$ respectively.

\begin{table*}
    \renewcommand{\arraystretch}{1.4}
	\centering
	\caption{Comparison of the luminosity-decay correlation results for the radio sample and the other wavebands. Col. (1): The waveband each sample belongs to. Col. (2): Number of GRBs in each sample. Cols. (3-4): Spearman's rank correlation coefficient and null hypothesis. Cols. (5-6): Partial Spearman's rank correlation coefficient and null hypothesis. Cols. (7-8): Linear regression analysis gradient and intercept.}
	\label{tab:summary}
	\begin{tabular}{lc c cc c cc c cc}
	\hline
    & & & \multicolumn{2}{|c|}{Spearman's rank} & & \multicolumn{2}{|c|}{Partial Spearman's rank} & & \multicolumn{2}{|c|}{Linear regression fit} \\
    \cline{4-5} 
    \cline{7-8}
    \cline{10-11}
	Sample & Number of GRBs & & Coefficient & Null hypothesis & & Coefficient & Null hypothesis & & Gradient & Intercept\\
    (1) & (2) & & (3) & (4) & & (5) & (6) & & (7) & (8) \\
	\hline
    Radio$^{a}$ & $16$ & & $-0.70 \pm 0.13$ & $2.58 \times 10^{-3}$ & & $-0.65$ & $8.90 \times 10^{-3}$ & & $-0.29^{+0.19}_{-0.28}$ & $8.12^{+8.86}_{-5.88}$\\
    GeV$^{b}$ & $13$ & & $-0.74 \pm 0.19$ & $4.11 \times 10^{-3}$ & & $-0.45$ & $1.37 \times 10^{-1}$ & & $-0.31^{+0.12}_{-0.09}$ & $14.43^{+4.55}_{-5.97}$\\
    Optical/UV$^{b}$ & $48$ & & $-0.58 \pm 0.11$ & $1.90 \times 10^{-5}$ & & $-0.50$ & $2.85 \times 10^{-4}$ & & $-0.28^{+0.04}_{-0.04}$ & $7.72^{+1.31}_{-1.31}$ \\
    X-ray$^{d}$ & $237$ & & $-0.59 \pm 0.09$ & $8.03 \times 10^{-8}$ & & $-0.63$ & $1.58 \times 10^{-6}$ & & $-0.27^{+0.04}_{-0.04}$ & $6.99^{+1.23}_{-1.11}$ \\
	\hline
	\end{tabular}
    \begin{minipage}[c]{2\columnwidth}
    {{\it Notes.} $^{a}$this paper (frequency: $8.5$ GHz), $^{b}$\citep{Hinds23}, $^{c}$\citep{oates12}, $^{d}$\citep{Racusin16}.}
 \end{minipage}
\end{table*}

\section{Discussion}
\label{sec:discussion}

We have shown that there is evidence for a luminosity-decay correlation in radio afterglows, suggesting that the brightest radio light curves decay on average more quickly than the faint ones, and that the correlation is not produced by chance or selection effects. In this section, we compare the correlation in different wavebands, we explore the potential causes of the correlation, and discuss the application of the correlation to other relevant studies.

\subsection{Comparison with the correlation in other wavebands}
\label{sec:comparison}

As mentioned in \S\ref{sec:introduction}, the luminosity-decay correlation has previously been observed in the optical/UV, X-ray and GeV wavebands and is presented in \cite{oates12,oates15,Racusin16,Hinds23}, for each waveband, respectively. 

First, we highlight and explain one immediate difference between the correlation in the radio and these other wavebands. The luminosity (and average rate of decay) were measured at (from) a rest frame time of 10 seconds in the GeV waveband, and 200 seconds in both the optical/UV and X-ray wavebands, whereas in the radio band these measurements are made at (from) a rest frame time of 10 days. The need to measure the parameters in the radio band at much later times compared to those in the other wavebands can be explained in the context of the standard afterglow model \citep[as described in \S\ref{sec:introduction}, and in][]{sar98}. In the standard afterglow model, the peak synchrotron frequency, $\nu_m$, decreases with time due to the hydrodynamical evolution of the relativistic shock as it collides with the external medium and decelerates. At early times, the radio band is initially in the regime of $\nu < \nu_m$ which results in the light curves rising, as observed. At later times, $\nu_m$ shifts to lower frequencies until eventually the radio band is in the regime of $\nu_m < \nu$ and the radio light curves start to decay. The optical, X-ray, and GeV bands are at higher frequencies than the radio band, so $\nu_m$ shifts below these frequency bands much earlier than it does for the radio band and consequently the light curves at these higher frequency bands start to decay much earlier. Thereby, we are required to make measurements at much later times in the radio band compared to the other wavebands in order to probe the spectral regime of $\nu_m < \nu$ when the light curves are decaying.

Table \ref{tab:summary} summarizes the results of the correlation analysis in the GeV, optical, and X-ray bands. First we compare the Spearman's rank results. The p-values in each waveband are all $\lesssim4\times10^{-3}$ suggesting that, at a significance level of $\sim3\sigma$, none of these correlations are observed due to chance. The correlation coefficient calculated in the radio is consistent with those found at other wavelengths within $1 \sigma$, suggesting that the strength of the correlation is similar in each waveband. Next, we compare the linear regression results. The gradients calculated in each waveband are consistent within $1\sigma$, suggesting that the behaviour of the relationship is similar in each waveband. Overall, this comparison suggests that the correlation is statistically significant, acts with similar strength, and has similar behaviour in different wavebands across the entire electromagnetic spectrum.

\subsection{Potential causes of the correlation}
\label{sec:causes}

\subsubsection{Constraints from the multiwavelength discovery}
\label{sec:constraints}
Our results show that the correlation is observed in multiple wavebands across the entire electromagnetic spectrum, covering 15 orders of magnitude in frequency from GeV photons to radio photons (see \S\ref{sec:comparison}). This supports the simple deduction that whatever mechanism is causing the correlation must be one which acts achromatically: with similar strength and behaviour in all wavebands. Therefore, we can discard any mechanism, that could produce the correlation only (or more strongly) in a narrow or select waveband. For example, the reverse shock, which is not expected to be the cause of the correlation following that it is expected to produce more emission in the optical and radio wavebands than in the X-ray, due to the lower Lorentz factors \citep{sar98,kobayashi99,kobayashi2000,laskar13,rho20,bri23}. This deduction was previously made and discussed in \cite{oates15,Racusin16,Hinds23} and is further supported here by the finding of the correlation in the radio band (extending further across the electromagnetic spectrum) with similar correlation strength and behaviour to that observed in the other bands. 

Although the reverse shock is not likely the cause of the correlation, it can be a prominent source of radiation; particularly at longer wavelengths such as the radio band. As a result, we consider the possible effect of the reverse shock on the correlation in this specific waveband. We searched the literature to determine if any GRBs in our sample have evidence of a significant reverse shock component in their radio data. Most of the GRBs in our final sample (all aside from 091020) have been individually studied in detail in the radio band. These studies indicate that only one GRB, 161219B, has firm evidence of a significant reverse shock component in its radio light curve \citep{las18} and an additional two GRBs, 000926 and 021004, have possible signatures, but lack firm evidence \citep{har01,kob03b}. Furthermore, in all three cases, the (possible) reverse shock component is no longer dominant by 10 days. Therefore, the reverse shock does not appear to be driving the correlation in our sample of radio light curves.

\subsubsection{Afterglow models}
The standard afterglow model is widely accepted as a means of explaining the production of multiwavelength GRB afterglows via synchrotron emission \citep{rees92, mes93, mes97, sar98}. In \cite{oates15}, Monte Carlo simulations of a basic synchrotron afterglow model were used to establish the expected relationship between the early time luminosity and the average rate of decay in optical/UV and X-ray afterglows. This model only considered emission from the forward shock and assumed a uniform medium, a uniform collimated jet, and no sustained energy injection. They found that the relationship expected between these parameters from this model is not consistent with the observed relationship, implying that this model is unsuccessful in explaining the cause of the observed luminosity-decay correlation at these frequencies. Instead, \cite{oates15} suggested that the correlation may be caused by a parameter/mechanism that was not previously considered in this basic afterglow model. As explained in \S\ref{sec:constraints}, this parameter/mechanism must be one which acts achromatically, such as time-varying microphysical parameters \citep[see;][]{mis21} or more complex circumburst density profiles, for example.

Alternatively, as afterglow emission is jetted, the correlation may be related to geometric effects such as the observer's viewing angle with respect to the jet axis. For example, afterglows that are observed off-axis, will be fainter, peak later, and have slower decline rates compared to their on-axis counterparts \citep{granot02,pan08}. Therefore, a range in observer viewing angle is a possible cause of the correlation since those afterglows which are initially brighter and decay more quickly on average may be attributed to GRBs which are viewed more on-axis. This was previously suggested as a possible cause in \cite{oates12}, and further discussed in \cite{oates15,Racusin16,Hinds23,gup24}. The results of the correlation observed in radio afterglows (see \S\ref{sec:results}) are consistent with this suggestion, of the correlation being caused by geometric effects. 

Structured outflows, in addition to viewing angles, will also affect the observed light curve properties as shown in Figure 3 of \cite{pan08}. This means that jet structure may add complexity to this scenario, possibly affecting the slope and strength of the correlation as well as the degree of scatter. 

If the correlation is due to the position of the observer relative to the jet, we would expect to see a correlation between the viewing angle and the luminosity-decay plane in each waveband. This will be investigated in Shilling et al. (in preparation). The correlation being due to viewing angle will also be explored using numerical simulations in Turnbull et al. (in preparation). Further tests to understand the cause of the correlation are detailed in \cite{oates12}.

\subsection{Comparison with different studies}

\subsubsection{Potentially different radio subclasses}
Recently, \cite{lloyd17,lloyd19} studied a sample of ${E_\mathrm{iso}} > 10^{52}$ erg GRBs. They split their sample of GRBs into radio-loud and radio-quiet categories depending on whether they had a radio detection or not. Based on the relationship between radio detectability and ${E_\mathrm{iso}}$, where radio detectability increases with ${E_\mathrm{iso}}$ \citep[as detailed in][]{Chandrafrail12}, they argued that their high ${E_\mathrm{iso}}$ GRBs without radio detections are more likely intrinsically radio-quiet instead of having radio afterglows that are simply undetected due to instrumentation sensitivity limits, and that contamination between the two categories is therefore reduced.

\cite{lloyd17,lloyd19} found that LGRBs that are radio-quiet have intrinsically shorter prompt durations than LGRBs that are radio-loud. On this basis, they suggest that radio-loud and radio-quiet LGRBs may be two distinct subpopulations entirely based on the presence or absence (where non-detections are not due to sensitivity limits) of radio emission. Similar conclusions are found in \cite{zha21} and, with a larger and updated sample, in \cite{chakra23}. A progenitor-based explanation for the difference in observed properties is presented in \cite{lloyd17,lloyd19}; such that, radio-loud LGRBs are produced by collapsars with a close binary companion and that radio-quiet LGRBs are produced by isolated field collapsars. The possibility of this progenitor-based explanation is discussed further in \cite{lloyd22}. 

Our sample of 16 GRBs does not contain any GRBs without radio afterglows, and therefore are exclusively radio-loud. Consequently, we cannot use this correlation to test whether radio-loud and radio-quiet GRBs are two distinct populations. We can only comment that radio-loud GRBs, at least those in our sample, do follow the luminosity-decay correlation, such that more intrinsically bright afterglows tend to decay on average more quickly. Examining the correlation in different wavebands, such as optical or X-ray, may be better suited for examining the possibility of radio-loud and -quiet GRBs belonging to two distinct populations. For instance, GRBs with X-ray afterglows could be split into two groups depending on if they are radio-loud or -quiet and analysed to test if the correlation is present, and consistent between, the two populations. This will be examined in Shilling et al. (in preparation).

\begin{center}
\begin{figure}
\includegraphics[width=\columnwidth]{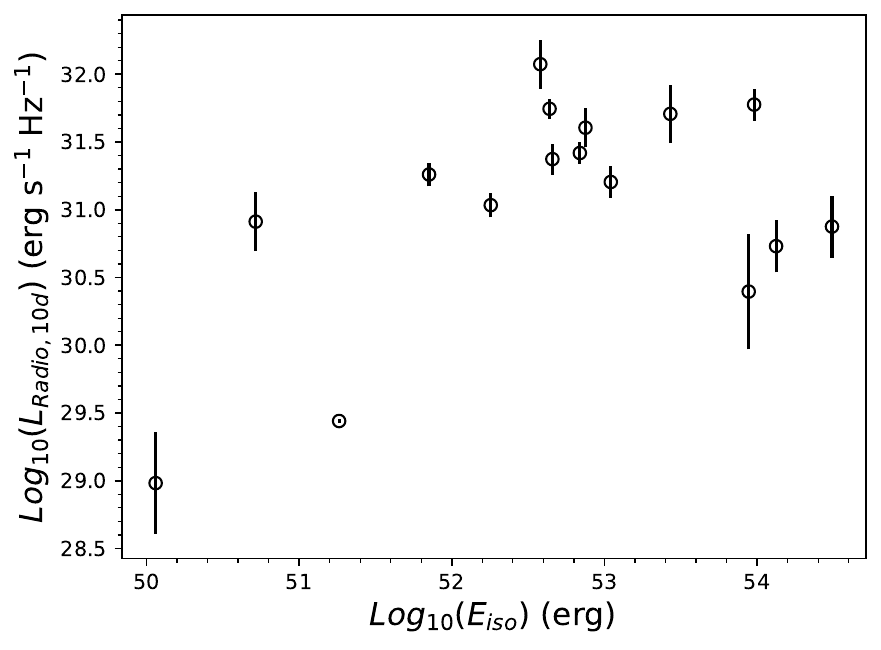}
    \caption{The 8.5 GHz luminosity measured at 10 days against the isotropic equivalent gamma-ray energies. All $E_\mathrm{iso}$ values are from \protect\cite{Chandrafrail12} apart from that of GRB 161219B, which is from \protect\cite{Ashall19}. There is a large degree of scatter and no statistically significant correlation is recovered with a Spearman's rank coefficient of 0.18 and a p-value of 0.51.}
    \label{fig:eiso}
\end{figure}
\end{center}

\subsubsection{Isotropic $\gamma$-ray energy and radio luminosity}
Motivated by the correlation between $E_\mathrm{iso}$ and luminosity at 200 seconds observed in the optical/UV and X-ray bands in \cite{oates15}, we examine this correlative property in the radio band. Using the final sample of radio afterglows in this paper, we test for a correlation between $E_\mathrm{iso}$ (see \S \ref{sec:eiso_dist_comparison}) and the luminosity measured at 10 days, as shown in Figure \ref{fig:eiso}. We calculate a Spearman's rank coefficient of 0.18 between these properties and a p-value of 0.51, which are insufficient to infer a statistically significant correlation.
Interestingly, this is contrary to the case in the optical/UV and X-ray bands where a statistically significant correlation is observed between $E_\mathrm{iso}$ and the luminosity at the respective earlier times in each band. The lack of an observed correlation between $E_\mathrm{iso}$ and $L_{\mathrm{Radio,10d}}$ in the radio is in agreement with the results in \cite{Chandrafrail12} (shown in their Figure 20), where no correlation is found between $E_\mathrm{iso}$ and $L_\mathrm{peak}$ in the radio band. However, the distribution of $E_\mathrm{iso}$ and $L_{\mathrm{Radio,10d}}$ shown in Figure \ref{fig:eiso} suggests that GRBs with larger $E_\mathrm{iso}$ tend to be more luminous in the radio band. This hints at a possible correlation, but a larger sample would be needed to confirm or rule out this possibility. 

\subsubsection{The plateau luminosity-time correlation}
GRB afterglows have been observed to have a feature in their light curve resembling a plateau, which are believed to be driven by a sustained period of continued energy injection from the GRB central engine \citep{nousek}. Previous studies have discovered a correlation in multiple wavebands between the luminosity, $L$, at the rest frame time at the end of the plateau, $T_a$, and $T_a$ itself. This correlation shows that afterglows that have a low luminosity at the end time of their plateau phase tend to have a plateau phase which ends at a later time. The $L(T_a)-T_a$ correlation was originally observed in X-ray afterglows and subsequently in optical and radio afterglows \citep{dai13,dai20,levine20}. There has been evidence to suggest that it may also exist in the GeV band \citep{dainotti21}.

The $L(T_a)-T_a$ correlation and the luminosity-decay correlation are likely connected since they are both observed in the afterglow phase, across multiple wavebands, and both use variables related to the intrinsic brightness. Although it is unclear how exactly they are connected, we infer that whatever is causing one correlation must also allow for the other. One benefit of the luminosity-decay correlation is that it allows for more GRBs to be included in samples as it is observed in those with and without plateaus \citep[see][]{Racusin16} so can be applied to all GRBs, provided they are well sampled, while the $L(T_a)-T_a$ plateau correlation requires a plateau and thereby can only be observed in the $\sim50\%$ of GRBs that have a plateau (in the X-ray band) \citep{gug24} in addition to being well sampled. 

\section{Conclusions}
\label{sec:conclusion}
For a sample of 16 selected GRB afterglow 8.5 GHz radio light curves, we measure the luminosity at 10 days in the rest frame, $L_{\mathrm{Radio,10d}}$, and the average rate of decay past this time, $\alpha_{>10d}$, using a simple power-law. We find evidence of a correlation between these parameters, suggesting that GRB radio afterglows which are initially bright tend to decay more quickly on average. The correlation has a Spearman's rank coefficient of $R_\mathrm{sp}=-0.70 \pm 0.13$ at a significance of $>3\sigma$ and a linear regression of $\alpha_{>10d} = -0.29^{+0.19}_{-0.28} \log \left(L_{\mathrm{Radio,10d}} \right) + 8.12^{+8.86}_{-5.88}$. We test if the correlation is produced by chance or selection criteria by using a Monte Carlo simulation, which suggests, at a confidence level of $>3\sigma$, that the correlation is indeed intrinsic and not produced by chance or selection criteria.

We discuss the possible causes of the correlation. The correlation coefficient and linear regression slope that we measure in the radio band are consistent with their counterparts measured in the optical/UV, X-ray and GeV wavebands in previous studies at $<1 \sigma$. This suggests that the potential mechanism(s) causing the correlation must produce a similar correlation strength, and behaviour, in each waveband. Given this constraint, one possible cause of the correlation is jet geometry; such that the correlation could be caused by different observer viewing angles within the jet. For example, larger viewing angles can result in fainter and slower decaying light curves compared to afterglows viewed more on-axis. Jet structure can affect observed light curve properties and therefore may also affect the correlation properties such as the Spearman's rank strength, linear regression slope, and scatter. An alternative possible cause of the correlation is a parameter/mechanism which regulates the energy release and the average rate of decay. This mechanism could cause the most luminous radio afterglows to lose their energy more quickly than less luminous afterglows. For instance, a time varying microphysical parameter - such as the electron energy distribution. Both of these scenarios, jet geometry and time varying microphysical parameter(s), should be investigated further as possible causes of the correlation.

To enhance the sample size of this study, we encourage future GRB radio follow-up observations to be taken at a common frequency for all GRBs. Also, to increase the number of data points for each light curve and enable more detailed radio studies, we encourage that radio follow-up observations of afterglows be taken at regular intervals over the entire lifetime of each afterglow, from the initial detection up until they fade below the instrument sensitivity limit.

\section*{Acknowledgements}
SPRS acknowledges support from an
STFC PhD studentship, the Faculty of Science and Technology at Lancaster University, and NASA under award number 80GSFC21M0002. MN is supported by the European Research Council (ERC) under the European Union’s Horizon 2020 research and innovation programme (grant agreement No.$\sim$948381). RG was sponsored by the National Aeronautics and Space Administration (NASA) through a contract with ORAU. The views and conclusions contained in this document are those of the authors and should not be interpreted as representing the official policies, either expressed or implied, of the National Aeronautics and Space Administration (NASA) or the U.S. Government. The U.S. Government is authorized to reproduce and distribute reprints for Government purposes notwithstanding any copyright notation herein. 

\section*{Data Availability}
The data were collected from the literature. A compiled catalogue of which can be made available upon request.

\bibliographystyle{mnras}
\bibliography{GRB} 

\bsp
\label{lastpage}
\end{document}